\documentclass[conference]{IEEEtran}
\ifCLASSINFOpdf
\else
\fi                                     

\usepackage{graphicx,graphics,epsfig,epstopdf,float}
\usepackage{amsmath} 
\usepackage{amssymb}  
\usepackage{mathrsfs}
\usepackage{enumerate}\usepackage{dsfont}
\usepackage{subfigure}
\usepackage{colortbl}
\usepackage{color}  
\usepackage{bm}
\usepackage{psfrag}
\usepackage{cite}
\usepackage{algorithm}
\usepackage{algorithmicx}
\usepackage{algpseudocode}
\usepackage{mdwlist}
\usepackage{longtable}
\usepackage{array}
\usepackage{acronym}  
\usepackage{url}

\usepackage{bm}
\usepackage{wrapfig}

\DeclareMathAlphabet{\mathsfbr}{OT1}{cmss}{m}{n}
\SetMathAlphabet{\mathsfbr}{bold}{OT1}{cmss}{bx}{n}
\DeclareRobustCommand{\msf}[1]{%
  \ifcat\noexpand#1\relax\msfgreek{#1}\else\mathsfbr{#1}\fi
}

\makeatletter
\newcommand{\msfgreek}[1]{\csname s\expandafter\@gobble\string#1\endcsname}
\makeatother

\DeclareFontEncoding{LGR}{}{} 
\DeclareSymbolFont{sfgreek}{LGR}{cmss}{m}{n}
\SetSymbolFont{sfgreek}{bold}{LGR}{cmss}{bx}{n}
\DeclareMathSymbol{\salpha}{\mathord}{sfgreek}{`a}
\DeclareMathSymbol{\sbeta}{\mathord}{sfgreek}{`b}
\DeclareMathSymbol{\sgamma}{\mathord}{sfgreek}{`g}
\DeclareMathSymbol{\sdelta}{\mathord}{sfgreek}{`d}
\DeclareMathSymbol{\sepsilon}{\mathord}{sfgreek}{`e}
\DeclareMathSymbol{\szeta}{\mathord}{sfgreek}{`z}
\DeclareMathSymbol{\seta}{\mathord}{sfgreek}{`h}
\DeclareMathSymbol{\stheta}{\mathord}{sfgreek}{`j}
\DeclareMathSymbol{\siota}{\mathord}{sfgreek}{`i}
\DeclareMathSymbol{\skappa}{\mathord}{sfgreek}{`k}
\DeclareMathSymbol{\slambda}{\mathord}{sfgreek}{`l}
\DeclareMathSymbol{\smu}{\mathord}{sfgreek}{`m}
\DeclareMathSymbol{\snu}{\mathord}{sfgreek}{`n}
\DeclareMathSymbol{\sxi}{\mathord}{sfgreek}{`x}
\DeclareMathSymbol{\somicron}{\mathord}{sfgreek}{`o}
\DeclareMathSymbol{\spi}{\mathord}{sfgreek}{`p}
\DeclareMathSymbol{\srho}{\mathord}{sfgreek}{`r}
\DeclareMathSymbol{\ssigma}{\mathord}{sfgreek}{`s}
\DeclareMathSymbol{\stau}{\mathord}{sfgreek}{`t}
\DeclareMathSymbol{\supsilon}{\mathord}{sfgreek}{`u}
\DeclareMathSymbol{\sphi}{\mathord}{sfgreek}{`f}
\DeclareMathSymbol{\schi}{\mathord}{sfgreek}{`q}
\DeclareMathSymbol{\spsi}{\mathord}{sfgreek}{`y}
\DeclareMathSymbol{\somega}{\mathord}{sfgreek}{`w}

\DeclareMathSymbol{\svarsigma}{\mathord}{sfgreek}{`c}

\DeclareMathSymbol{\sGamma}{\mathalpha}{sfgreek}{`G}
\DeclareMathSymbol{\sDelta}{\mathalpha}{sfgreek}{`D}
\DeclareMathSymbol{\sTheta}{\mathalpha}{sfgreek}{`J}
\DeclareMathSymbol{\sLambda}{\mathalpha}{sfgreek}{`L}
\DeclareMathSymbol{\sXi}{\mathalpha}{sfgreek}{`X}
\DeclareMathSymbol{\sPi}{\mathalpha}{sfgreek}{`P}
\DeclareMathSymbol{\sSigma}{\mathalpha}{sfgreek}{`S}
\DeclareMathSymbol{\sUpsilon}{\mathalpha}{sfgreek}{`U}
\DeclareMathSymbol{\sPhi}{\mathalpha}{sfgreek}{`F}
\DeclareMathSymbol{\sPsi}{\mathalpha}{sfgreek}{`Y}
\DeclareMathSymbol{\sOmega}{\mathalpha}{sfgreek}{`W}

\DeclareRobustCommand{\mcal}[1]{%
  \ifcat\noexpand#1\relax\mathnormal{#1}\else\cal{#1}\fi
}
\DeclareRobustCommand{\BM}[1]{%
  \ifcat\noexpand#1\relax\bm{\boldUppercaseItalicGreek{#1}}\else\bm{#1}\fi
}
\makeatletter
\newcommand{\boldUppercaseItalicGreek}[1]{\csname var\expandafter\@gobble\string#1\endcsname}
\makeatother
\newcommand{\rv}[1]{\MakeLowercase{\msf{#1}}} 
\newcommand{\RV}[1]{\bm{\MakeLowercase{\msf{#1}}}}  

\newcommand{\V}[1]{\bm{#1}} 
\newcommand{\M}[1]{\BM{#1}} 
\newcommand{\Set}[1]{\mcal{#1}} 

\definecolor{BLUE}{rgb}{0,0,1}
\acrodef{gnss}[GNSS]{global navigation satellite system}
\acrodef{rf}[RF]{radio frequency}
\acrodef{aoa}[AOA]{angle-of-arrival}
\acrodef{rss}[RSS]{received signal strength}
\acrodef{toa}[TOA]{time-of-arrival}
\acrodef{tdoa}[TDOA]{time-difference-of-arrival}
\acrodef{rtt}[RTT]{round-trip time}
\acrodef{fdd}[FDD]{frequency division duplex}
\acrodef{tdd}[TDD]{time division duplex}
\acrodef{fd}[FD]{full-duplex}
\acrodef{sdp}[SDP]{semidefinite programming}
\acrodef{crlb}[CRLB]{Cram\'{e}r-Rao lower bound}
\acrodef{nc}[NC]{narrow correlator}
\acrodef{sc}[SC]{storbe correlator}
\acrodef{pll}[PLL]{phase locked loop}
\acrodef{mp}[MP]{multipath}
\acrodef{sp}[SP]{single path}
\acrodef{ff}[FF]{flat fading}
\acrodef{mds}[MDS]{multidimensional scaling}
\acrodef{snr}[SNR]{signal-to-noise ratio}
\acrodef{los}[LOS]{line-of-sight}
\acrodef{nlos}[NLOS]{non-line-of-sight}
\acrodef{sic}[SIC]{serial interference cancelation}
\acrodef{pic}[PIC]{parallel interference cancelation}
\acrodef{adc}[ADC]{analog-to-digital converter}
\acrodef{bp}[BP]{basis pursuit}
\acrodef{lasso}[LASSO]{least absolute shrinkage and selection operator}
\acrodef{omp}[OMP]{orthogonal matching pursuit}
\acrodef{lls}[LLS]{linear least squares}
\acrodef{wlls}[WLLS]{weighted linear least squares}
\acrodef{nlls}[NLLS]{nonlinear least squares}
\acrodef{awgn}[AWGN]{additive white Gaussian noise}
\acrodef{cirf}[CIRF]{channel impulse response function}
\acrodef{irf}[IRF]{impulse response function}
\acrodef{llr}[LLR]{log-likelihood ratio}
\acrodef{llrs}[LLRs]{log-likelihood ratios}
\acrodef{fim}[FIM]{Fisher information matrix}
\acrodef{efim}[EFIM]{equivalent Fisher information matrix}
\acrodef{mse}[MSE]{mean squared error}
\acrodef{peb}[PEB]{position error bound}
\acrodef{rmse}[RMSE]{root mean squared error}
\acrodef{seb}[SEB]{synchronization error bound}
\acrodef{imu}[IMU]{inertial measurement unit}
\acrodef{nls}[NLS]{network localization and synchronization}
\acrodef{Nls}[NLS]{Network localization and synchronization}
\acrodef{reb}[REB]{ranging error bound}
\acrodef{co}[CO]{clock offset}
\acrodef{cdma}[CDMA]{code-division multiple-access}
\acrodef{pdf}[PDF]{probability density function}%

\title{\LARGE \bf
 Relative Distributed Formation and Obstacle Avoidance with  Multi-agent Reinforcement Learning
}
 
\author{Yuzi Yan, Xiaoxiang Li, Xinyou Qiu, Jiantao Qiu, Jian Wang, Yu Wang, Yuan Shen 
\thanks{The authors are with the Department of Electronic Engineering, Tsinghua University, Beijing 100084, China, and also with the Beijing National Research Center for Information Science and Technology, Beijing 100084, China. E-mail: \{yanyuzi21, lxx17, qxy18, qjt15\}@mails.tsinghua.edu.cn; \{jian-wang, yu-wang, shenyuan\_ee\}@tsinghua.edu.cn.}%
}

\begin{document}

\maketitle
\thispagestyle{empty}
\pagestyle{empty}

\begin{abstract}

Multi-agent formation as well as obstacle avoidance is one of the most actively studied topics in the field of multi-agent systems. Although some classic controllers like model predictive control (MPC) and fuzzy control achieve a certain measure of success, most of them require precise global information which is not accessible in harsh environments. On the other hand, some reinforcement learning (RL) based 
approaches adopt the leader-follower structure to organize different agents' behaviors, which sacrifices the collaboration between agents thus suffering from bottlenecks in maneuverability and robustness. In this paper, we propose a distributed formation and obstacle avoidance method based on multi-agent reinforcement learning (MARL). Agents in our system only utilize local and relative information to make decisions and control themselves distributively. Agent in the multi-agent system will reorganize 
themselves into a new topology quickly in case that any of them is disconnected. Our method achieves better performance regarding formation error, formation convergence rate and on-par success rate of obstacle avoidance compared with baselines (both classic control methods and another RL-based method). The feasibility of our method is verified by both  simulation and hardware implementation with Ackermann-steering vehicles\footnote{Simulation and hardware implementation demos can be found at \url{https://sgroupresearch.github.io/relativeformation/}.}. 

\end{abstract}

\section{INTRODUCTION}
Formation control while avoiding obstacles is one of the most basic function of an multi-agent system (MAS). In scenarios like Internet of Vehicles, the autonomous platooning (as a formation task) and overtaking (as an obstacle avoidance task) are the most common and important maneuvers.

Most previous studies~\cite{oh2015survey,sun2017distributed,lin2013leader,ogren2003obstacle,de2008dynamic,wen2017formation} regard the whole task as an optimization problem to plan the agents' route and movement according to the destination and reward function, while under constraints like avoiding obstacles and other agents during the motion. Since the optimization problem tends to be non-convex, some related works based on classic hierarchical control like \emph{model predictive control (MPC)}~\cite{LiMaShe:C19,dai2017distributed} or \emph{fuzzy control}~\cite{naranjo2008lane} are proposed to deal with the problem. However, most of them require high-precision global information like GPS and digital maps, leading to inapplicability in harsh environments such as search-and-rescue in emergency disasters. Besides, the collaboration between agents are not fully considered in these traditional methods, which means there is still a lot of room for improvement in terms of multi-agent collaboration. 

During the past few decades, the maturity of intelligent agent has been largely enhanced thanks to the deep combination of reinforcement learning (RL) and control theory~\cite{iima2015swarm,sui2020formation,kobayashi2003re,sui2018path}. Some previous works~\cite{wen2017optimized, knopp2017formation} conduct RL to realize automatic formation control and obstacle avoidance. But most of them fail to get rid of the leader-follower structure, and mainly focus on controlling a certain single agent~\cite{zhou2019adaptive,yang2006multiagent,miah2020model,qiu2021drl,Han2019}. If the leader agent is destroyed or disconnected, the whole system will collapse. What's more, common RL-based works are verified only by numerical simulation. A few works implement their algorithms on hardware platforms but only take the omnidirectional wheel model into consideration~\cite{RenChao2014,Giovanni2009 }, which is not enough to be used in practical Ackermann-steering\footnote{Ackermann-steering geometry is the practical model as human driven vehicles. Compared with other omnidirectional wheel systems, it has more constraints and faces challenges in convergence and time consumed for training.}~\cite{mitchell2006analysis} vehicular system.

\begin{figure}[t]
  \centering
  \includegraphics[width=1.0\linewidth]{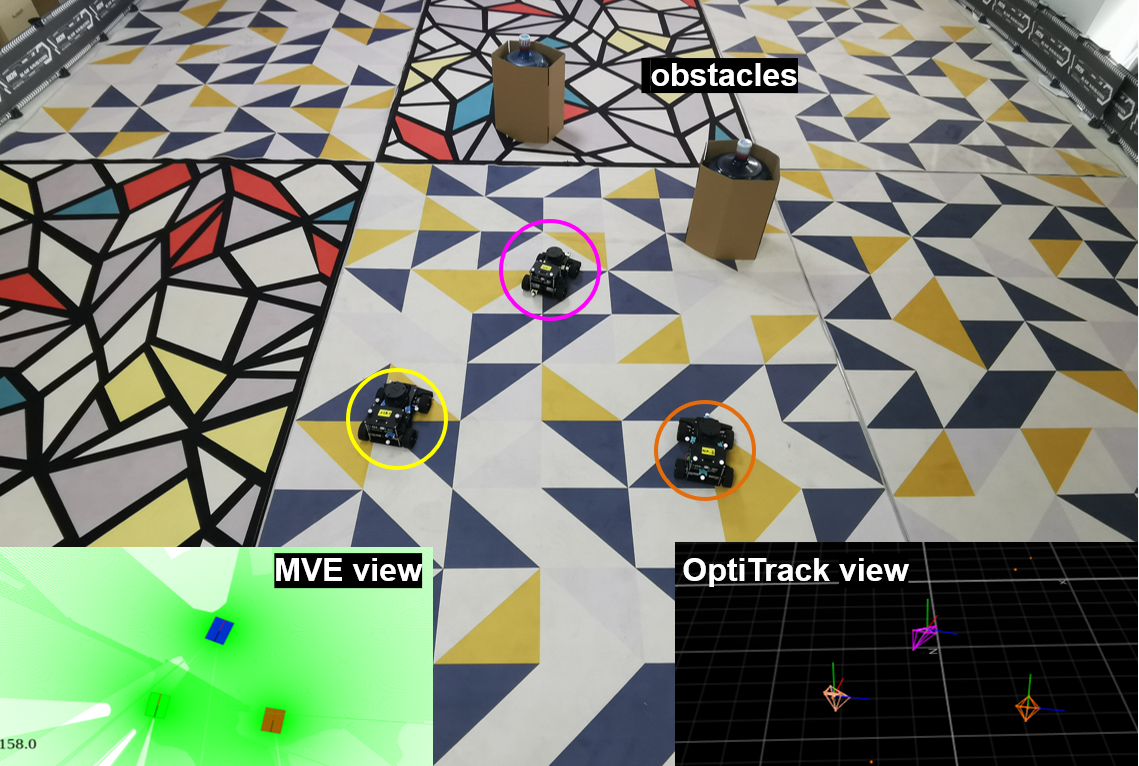}\\
  \centering
  \caption{The scenario of our problem. Ackermann-steering agents are placed randomly and meant to complete a formation task while avoiding obstacles detected by Lidar sensors. The sub-picture in the bottom left shows the results obtained by our self-developed simulator, called Multi-Vehicle Environment(MVE). The sub-picture in the lower right shows the precise coordinates of agents obtained by OptiTrack.}\label{fig:scenario}
\end{figure}

  
In this paper, we design a distributed formation and obstacles avoidance algorithm with Multi-Agent Proximal Policy Optimization (MAPPO)~\cite{schulman2017proximal,yu2021surprising}. The proposed method requires only relative information which is easily available for real systems, and the control policy can be executed distributively. The \textbf{contributions} of our work are as follows: 1) Distributed. We put forward a relative formation strategy that is independent of global position information. Agents adjust their postures by taking into account the network topology obtained by spacial-temporal cooperation rather than absolute coordinates or orientation angle information. We avoid the leader-follower structure and train a policy that supports decentralized execution, which is more robust. 2) Adaptive. Multiple formation strategies are integrated in our model through policy distillation. If any agent in the formation is destroyed or disconnected, other agents will reorganize themselves into a new topology adaptively to continue their work. Besides, we use curriculum learning~\cite{narvekar2020curriculum} to accelerate training regarding obstacle avoidance, improving the stability of the MAS in a complex environment.
3) Effective. We compare our method with existing traditional control algorithms (MPC~\cite{LiMaShe:C19}, fuzzy control~\cite{naranjo2008lane}) and a RL-based leader-follower method~\cite{qiu2021drl} by numerical simulations. The results of average formation error, formation convergence rate and success rate of obstacle avoidance illustrate that our method achieves better performance than the baselines.
4) Practical.  We conduct our method on a hardware platform using intelligent cars with Ackermann-steering geometry as agents.


\emph{Notation}: Throughout this paper, variables, vectors, and matrices are written as italic letters $x$, bold italic letters $\V{x}$, and bold capital italic letters $\M{X}$, respectively. Random variables and random vectors are written as sans serif letter $\rv{x}$ and bold letters $\RV{x}$, respectively. The notation $\mathbb{E}_{\RV{x}} \{\cdot \}$ is the expectation operator with respect to the random vector $\RV{x}$, and $\mathds{1}(\cdot)$ is the indicator function which equals $1$ if the condition is true and equals $0$ otherwise.

\section{PROBLEM FORMULATION}

\subsection{Relative Localization and Formation Error}
High-precision location information is a prerequisite and important guarantee for complex tasks such as formation and obstacle avoidance. 
The state of the art studies mainly focus on the global localization optimization which is high-cost and unguaranteed in harsh environments\cite{SheWymWin:J10}.  In scenarios like Internet of Vehicles, people pay more attention to relative relationships which reflect the shape of network geometry\cite{LiuWanSheWanShe:J21}, since the relative topology is sufficient to complete maneuvers like overtaking or formation and much easier to be obtained.

Considering a two-dimensional formation that consists of $N$ agents and the set of agents is denoted as $\mathcal{N}$.
The global position of agent $i$ is denoted as $ \V{p}_i = [ x_i \  y_i ]^{\rm{T}}$. 
In the local coordinate system of agent $i$, the relative position of  any  other  agent $j$ is denoted  as $\V{{p}}_{i\leftarrow j}= [x_{j} -x_{i}\  y_{j}-y_{i}]^{\rm{T}}$. The relative position parameter vector of the  formation  is  denoted  as ${\V{p}}_{ i \leftarrow \tilde{i}}= [\V{ {p}}_{i\leftarrow 1} \ldots \V{ {p}}_{i\leftarrow N}]^{\rm{T}}$. For a given formation positioned as $ \V{p} $, the \emph{equivalent geometry} of this formation is denoted as 
\begin{equation}\label{sec:formerr}
T_{\V{\omega}}(\V{p})=\left(\M{I}_{N} \otimes  \boldsymbol{\Gamma}_{{\vartheta }} \right)   \V{p}+\V{1}_{N} \otimes  \V{t}, 
\end{equation}
where $ \V{\omega} = [\boldsymbol{\Gamma}_{{\vartheta }} \  \V{t} ] $, $\boldsymbol{\Gamma}_{{\vartheta }}$ denotes the rotation matrix of angle $\vartheta  \in[0,2 \pi)$, and $ \V{t} \in  \mathbb{R}^{2}$ denotes the translation in $x,\ y$ axes. 

Since two formations are considered {equivalent} if one can be transformed into another through rigid body transformation like translation and rotation. The formation error $\mathcal{E}(\V{p},\V{q})$  between two given formations $\V{p}$  and  $\V{q}$ is defined as the the squared Euclidean distance between the equivalent geometry sets $T_{\V{\omega}}(\V{p})$   and $T_{\V{\omega}}(\V{q})$:
\begin{equation}
\begin{aligned}\label{formula:formation_error}
\mathcal{E}(\V{p},\V{q})  &= \min_{\V{p}\in T_{\V{\omega}}(\V{p}),\V{q} \in T_{\V{\omega}}(\V{q})} || \V{p} -\V{q}||^2 \\
&= \min_{\V{\omega}^*}{ {\|\V{p} - T_{\boldsymbol{\omega}^*}(\V{q}) \|^2} } 
\end{aligned}    
\end{equation}

\emph{Remark 1}: It's obvious that $T_{\V{\omega}}({\V{p}}_{i\leftarrow \tilde{i} }) = T_{\V{\omega}}({\V{p}}_{j\leftarrow \tilde{j} })$ for any two agents $i$ and $j$ in the same formation topology. To keep the
notation simple, we use $\breve{\RV{{p}}} $ to represent the relative topology obtained by agent $i, \forall i\in \mathcal {N}$.  

\subsection{Multi-agent Reinforcement Learning}
The relative distributed formation as well as obstacle avoidance can be regarded as a fully-cooperative problem, which is solved under a MARL framework. 

The MARL process of $N$ agents can be modeled as an extension of $N$ Markov decision processes\cite{lowe2017multi}. It composes of a state space $\Set{S}$ describing the possible configurations of all agents, a set of actions actions ${\Set{A}}_1,{\Set{A}}_2,...,{\Set{A}}_N $ and a set of observations  ${\Set{O}}_1,{\Set{O}}_2,...,{\Set{O}}_N $. 

Following~\cite{qiu2021drl}, the action output of each agent $i, i \in \mathcal{N} $ includes 4 control variable: $[\omega_L, \omega_R, v_F, v_B] \in {\Set{A}}_i$, where $\omega_L$ and $\omega_R$ represent the angular velocity of turning left and right; $v_F$ and $v_B$ represent the speed of moving forward and backward, respectively. 

As for the observations, ${\Set{O}}_i$ consists of variables as follow:
1) the relative distance $\rv{d}_{ij}$ and angle $\rv{\theta}_{ij}$ towards other agent $j, j \in \mathcal{N}$, by which we estimate the network topology and calculate the formation error $\mathcal{E}$;  
2) the relative distance $D_i$ and angle $\Theta_i$ towards the destination; 
3) the shortest distance of nearby obstacles detected by the Lidar sensor $\rv{d}_{im} = \min \RV{d}_i$ and its corresponding direction $\rv{\theta}_{im}$ according to the agent's coordinate, where $\RV{d}_i=[\rv{d}_{i1}\ \rv{d}_{i2} \ldots \rv{d}_{iM}]^{\rm{T}}$ is the detected result with angle resolution $2\pi/M$. 

Each agent $i$ obtains their own \emph{reward}   $r_i(\V{s}^t ,\V{a}_i^t ):{\Set{S}} \times {\Set{A}}_{i} \rightarrow \mathbb{R}$ to measure the feedback cost when taking action $\V{a}^t_i
 $ with state $\V{s}^t $ at time slot $t$, where $\V{a}^t_i$ and $\V{s}^t$ are always omitted for the simplicity of notion.  A \emph{policy}, denoted as $\V{a}_i \sim \V{\pi}_{{\theta}_i}( \cdot | s): \mathcal{S} \rightarrow \mathcal{P}(\mathcal{A}_i)
$, projects states to the probability measures on $\mathcal{A}_i$ which returns the probability
density of available state and action pairs $(\V{s}, {\V{a}_i})$. ${\theta}_i$ refers to parameters of the function $\V{\pi}$, and is always omitted for the simplicity.

RL involves estimating the total expected reward $\eta({\V{\pi}}) =  \mathbb{E}_{\rv{\pi}} \left[\sum_{{t}=0}^{\infty} \gamma^{t} r (\V{s}^t, \V{a}^{t})  \right]  $ with policy gradient methods (e.g. PPO \cite{schulman2017proximal}), who optimizes the parameter ${\theta}$ of the policy according to the explored data.

Among all policy gradient methods, PPO shows its efficiency in both stabilizing the policy and exploring for optimal results. Moreover, it has become the most powerful baseline algorithm of DRL due to its generalization to various tasks, including MAS. In rest of the paper, our method applies MAPPO~\cite{yu2021surprising}, an advanced version of PPO for multi-agent tasks to solve the formation problem.

\section{APPROACH}

\subsection{Reward Function Design}

One of the most important tasks of RL is the design of reward function. The proposed reward function is divided into three parts: relative formation reward, navigation reward and obstacle avoidance reward 
\begin{equation}\label{reward3}
r = r_{\text{form}} + \alpha r_\text{navi} + \beta r_\text{avoid},
\end{equation}
where $\alpha$ and $\beta$ are the hyper-parameters for reward trade-off.

The relative formation reward is designed based on the formation error as defined in (\ref{formula:formation_error}).
Given an ideal formation topology $\V{q}$, the formation error $\mathcal{E}(\breve{\RV{p}}, {\V{q}})$ can be used to measure the difference between the ideal formation and the actual formation, which is an optimizable goal in the RL framework. In the proposed method, the relative formation reward is defined as:
\begin{equation}\label{reward_form}
    r_{\text{form}}=-\frac{\mathcal{E}( {\breve{\RV{{p}}}} ,\V{q})}{G(\V{q})}.
\end{equation}
 $G(\V{q}) =\max \limits_{i,j\in \mathcal{N}} d_{ij}^2$ is a normalization factor related only to the size and topology of the ideal formation, where $d_{ij}$ represents the distance between agent $i$ and agent $j$ in the formation. It is introduced  to  normalize  the  reward  regardless  of different formation scales, which avoid repeated adjustment of the hyper-parameters $\alpha$ and $\beta$.

  The navigation reward is designed to reflect the efficiency of the agent moving towards the destination. Following~\cite{WanWanSheZha:J19}, the navigation reward is designed as $r_\text{navi}= D_{i}^{t-1}-D_{i}^{t}$, where $d_{i}^{t}$ represents the distance from agent $i$ to the destination at time slot $t$. 
  
The obstacle avoidance reward is designed to reflect the success rate of collision avoidance. Following~\cite{jin2019efficient}, the number of collisions in time slot $t$ is counted:

\begin{align}
R_\text{avoid}= -\sum_{ i \in \mathcal{N}, \  j \in \mathcal{N} \cup\mathcal{M} } \mathds{1}(d_{ij} < \delta_{ij} )   \notag    ,
\end{align}
 where $\delta_{ij}$ is the collision margin of agent $i$ with entity $j$, where $\mathcal{N}$ and $\mathcal{M}$ denotes the set of agents and obstacles, respectively. Since Lidar or ultrasonic sensors can not distinguish whether the detected object is an agent or an obstacle, we treat all entities equally to be avoided.
 
\subsection{Policy Distillation for Formation Adaptation}

We call it \emph{formation adaptation} that agents need to reorganize themselves into a new topology in case that any agent is disconnected. To achieve formation adaptation, we conduct policy distillation~\cite{green2019distillation}, a method that integrate the learned policies to handle different numbers of agents to complete the formation. 
For example, if one agent in a regular pentagonal formation completed by 5 agents is damaged due to collision of obstacles, the remaining 4 agents need to reorganize themselves into a square to move on.

The maximum number of agents in the formation is set to be $N$, and the corresponding ideal formation topology as $\V{q}_{N}$. For $n$ agents participating in the formation ($n < N$), we pre-set its corresponding ideal formation topology as $\V{q}_n$. In order to allow different numbers of agents to complete the formation adaptively, we train teacher models according to different $n$ and $\V{q}_n$. Then we use policy distillation to train a student model that can handle multiple situations from the teacher models.

\begin{figure}[thpb]
  \centering
  \includegraphics[width=1.0\linewidth]{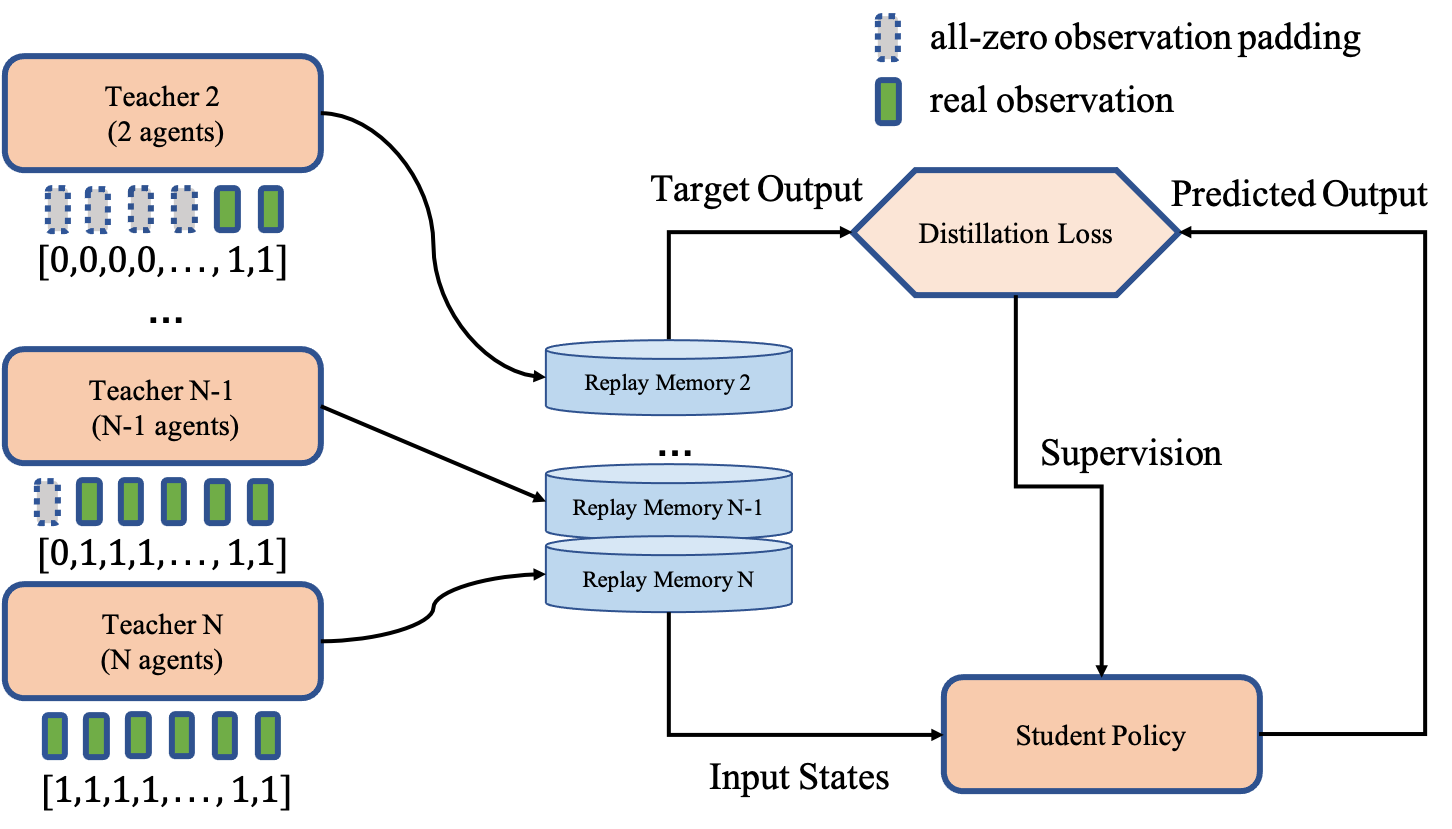}
  \caption{Policy distillation for formation adaptation. The observations (grey blocks for all-zero observation padding and green blocks for real observation) as well as corresponding death masking are input into teacher models for training. Teacher models generate input states and target output for supervised learning of student model where KL divergence is used as loss function.}
  \label{fig:distillation}
\end{figure}

We train several teacher models which share the same structure but with different $n$ and $\V{q}_n$. Connected agents are agents participating in the formation. We set \emph{death masking}~\cite{yu2021surprising} to mark the status of agents. The corresponding value of a connected agent in the \emph{death masking} is set to be $1$ while that of a disconnected agent is set to be $0$. Due to the different number of agents participating in the formation, the length of the input observations (such as the relative distance and relative angle of other agents) is different for different teacher models. To align the length of observation inputs, all-zero observation padding caused by \emph{death masking} is set for disconnected agents as shown in Fig.~\ref{fig:distillation}. 

In the stage of policy distillation, the trained teacher models produce inputs and targets, which are then stored in separate memory buffers. The student model learns from the data stored sequentially, switching to a different memory buffer every episode, just as in~\cite{rusu2015policy}.  We adopt the distillation setup of~\cite{green2019distillation} and use the KL divergence as the loss function:

$$L(p_T(s)|p_S(s))=\sum\limits_{i=1}^{\lvert \Set{A} \rvert}p_{k,T}(s)\ln{\frac{ p_{k,T}(s)}{ p_{k,S}(s)}},$$
where $\lvert \Set{A} \rvert$ is the dimension of the action space, $p_{k,S}(s)$ and $p_{k,T}(s)$ represent the probabilities for action $k$ of the student model $S$ and the teacher model $T$ given state observation $s$, respectively.

\begin{figure}[thpb]
  \centering
  \includegraphics[width=0.7\linewidth]{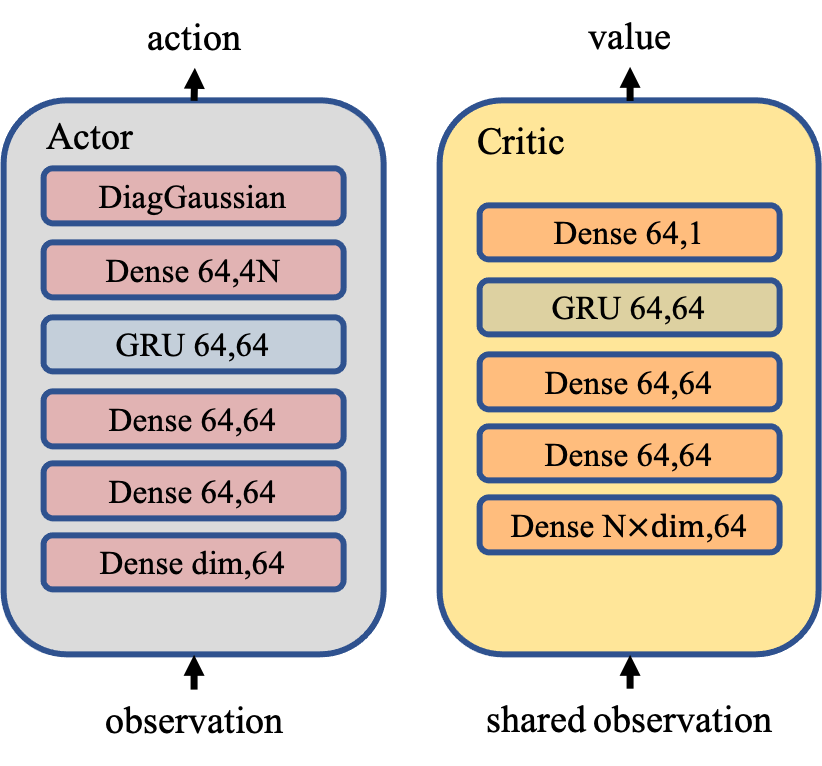}
  \caption{The model structure of actor-critic networks we use.  Note that we also use tricks such as observation normalization, layer normalization, ReLU activation, etc., just as in~\cite{yu2021surprising}, but do not show here mainly for simplicity.}
  \label{fig:network}
\end{figure}

\subsection{Curriculum Learning for Obstacle Avoidance}
As mentioned in Section \uppercase\expandafter{\romannumeral2}, the optimization problem tends to be non-convex. Training the network directly in a complex environment is likely to cause non-convergence.
In order to reduce the difficulty of training and speed up convergence, we use the idea of \emph{predefined curriculum learning}~\cite{bengio2009curriculum} and set different levels of difficulties for obstacle avoidance.

The specific curriculum settings are as follows:
We set a total of 5 difficulty levels according to the density of obstacles. At the  beginning, no obstacle will be placed in the environment (level-0). The agents focus mainly on learning basic formation strategies and navigating themselves to the destination. As the reward curve gradually converges, the level of difficulty will increase and more obstacles of different sizes will gradually appear in the environment. The agents are able to learn how to avoid obstacles to reach the destination as well as maintaining the ideal formation as stably as possible. The settings and results will be introduced in detail in the next section.

\begin{figure*}[thpb]
  \centering
  \includegraphics[width=1.0\linewidth]{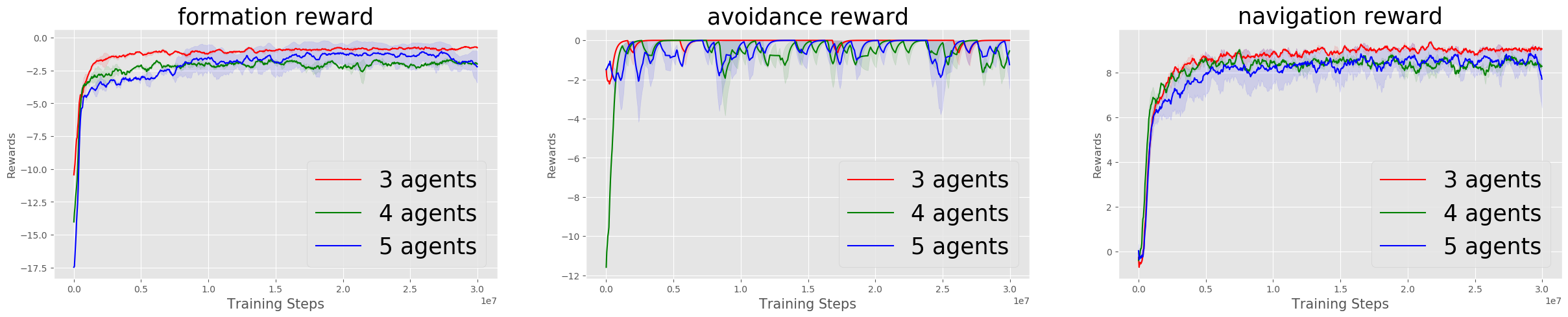}
  \caption{The curve of the reward function (formation, obstacle avoidance and navigation) when different numbers of agents are in formation. Due to normalization, the trade-off parameters $\alpha$ and $\beta$ defined in (\ref{reward3}) do not need to be adjusted manually as the number of agents changes.}
  \label{fig:train}
\end{figure*}

\section{EXPERIMENTS AND RESULTS}

To validate the effectiveness of our algorithm, we compare our method with several baseline methods including MPC~\cite{LiMaShe:C19}, fuzzy control~\cite{naranjo2008lane}, and a RL-based leader-follower scheme~\cite{qiu2021drl} on the Multi-agent Particle-world Environment (MPE)~\cite{lowe2017multi}.
To further evaluate the practicability of our method under physical constraints, we not only develop a new simulator, called Multi-Vehicle Environment (MVE)~\footnote{\url{https://github.com/efc-robot/MultiVehicleEnv}}, which supports the Ackermann-steering model rather than omnidirectional wheel model, but also implement our algorithm on a corresponding hardware platform. 

\subsection{Model Configuration}

In our model, the actor network and the critic network both consist of 3 dense layers, 1 GRU layer and 1 dense layer in order, as shown in Fig.~\ref{fig:network}. The hidden sizes of the dense layer and GRU layer are set to be 64. We follow the common practices in PPO implementation, including Generalized Advantage Estimation (GAE) ~\cite{schulman2015high} with advantage normalization, observation normalization, gradient clipping, layer normalization, ReLU activation with orthogonal initialization. Following the PopArt technique proposed by~\cite{hessel2019multi}, we normalize the values by a running average over the value estimates to stabilize value learning.



Following the {centralized training decentralized execution} principle\cite{chen2019new}, the shared observation (i.e. observations from all agents) is input to critic network. In training process,  we take 30,000,000 steps to optimize our model in total. We train our model on 1 NVIDIA RTX3090 GPU and AMD EPYC 7R32 48-Core CPU. The Adam optimizer\cite{kingma2014adam} is used with $\beta_1=0.9$, $\beta_2=0.98$, $\epsilon=10^{-3}$.

\emph{Remark 2}:
The dynamics design of the vehicle takes into account the characteristics of the Ackerman-steering model. In the MPE environment, the model is simplified by setting wheelbase to $0$. In the MVE environment, we designed the vehicular body based on the actual hardware, which will be described in detail at Section~\ref{MVE}.

\subsection{Simulation Results in MPE}

\begin{figure}[thpb]
  \centering
  \includegraphics[width=1.0\linewidth]{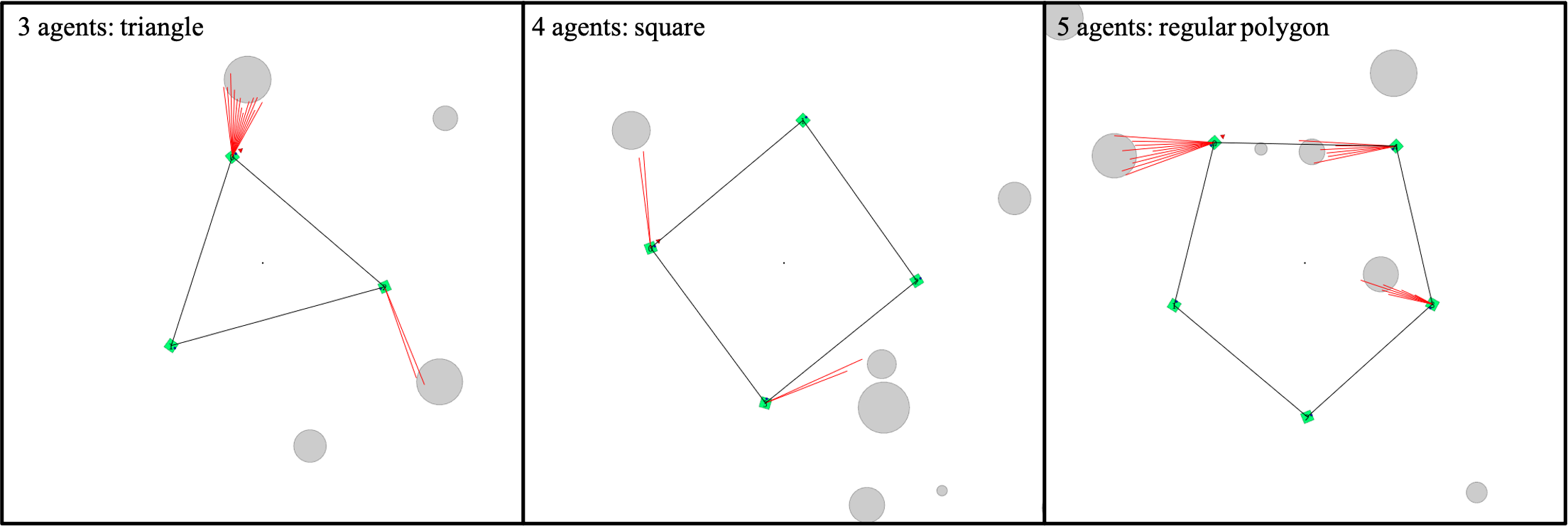}
  \caption{Simulation results in Multi-agent Particle-world Environment (MPE). The agents (green squares) are navigating towards the destination that is out of the screen, as well as maintaining the formation and avoiding obstacles (grey circles). The red rays represent the field of view of the Lidar equipped by agents. }
  \label{fig:MPE1}
\end{figure}

The scenario settings are as follows:
The maximum speed of agents is restricted to be $1 $m/s. 
Agents are squares with a side length of $0.01\text{m}$. Agents are randomly initiated in the range of $x\in[-2\text{m},2\text{m}],\ y\in[-2\text{m},2\text{m}]$ and the destination will be initiated at a distance of at least $36m$ from the agent's initial location. 
Obstacles are circles with random radius between $[0.01\text{m}, 0.05\text{m}]$. Obstacles are placed with uniform distribution between the start point $[0\text{m},0\text{m}]$ and the destination.
The amount of the obstacles randomly distributed on the map will increase according to different difficulty levels. The rendering of our scenario in MPE is shown in Fig.~\ref{fig:MPE1}

\subsubsection{Training Process}

As is shown in Fig.~\ref{fig:train}, we train the corresponding models for the formation with 3, 4 and 5 agents, respectively, where the ideal formation topology is set to be regular polygon. 
It is demonstrated that in the later stage of training, our model is able to organize and maintain formation finely (the formation error rises close to $0$) while avoiding the obstacles and advancing to the destination.

\subsubsection{Baseline Comparison}

We compare our proposed method with other methods, including: 1) MPC~\cite{LiMaShe:C19}, 2) fuzzy control~\cite{naranjo2008lane}, 3) RL-based leader-follower scheme~\cite{qiu2021drl}. In baseline comparison, the agent number is set to be 4, and the difficulty level is set to be level-3. The ideal formation topology is set to be a square with side length of 4m.

The traditional controllers (MPC and fuzzy control) treat the multi-task scenario as a motion-planning problem. The desired position of each sub-task will be taken as input to the controller. To ensure the performance of compound behaviors, the sub-tasks are assigned to different priorities with fine-tuned thresholds. The controller will execute the task with the highest priority at a time step. Since safety is the first concern, the avoiding behavior is the top priority among all, while formation behavior comes second. Thus, the navigation behavior won’t be triggered unless the agents enter a safety area and the formation error converges. The obstacle-avoidance behavior is designed with a stream-based path planner~\cite{WanCheFanMa:J14}. 

The RL-based leader-follower scheme uses the same network architecture as our method to train the agents to form the ideal topology by tracking a virtual leader. Each agent is only responsible of tracking its relative position towards the leader and obstacles, as demonstrated in~\cite{qiu2021drl}.

\begin{figure}[thpb]
  \centering
  \includegraphics[width=1.0\linewidth]{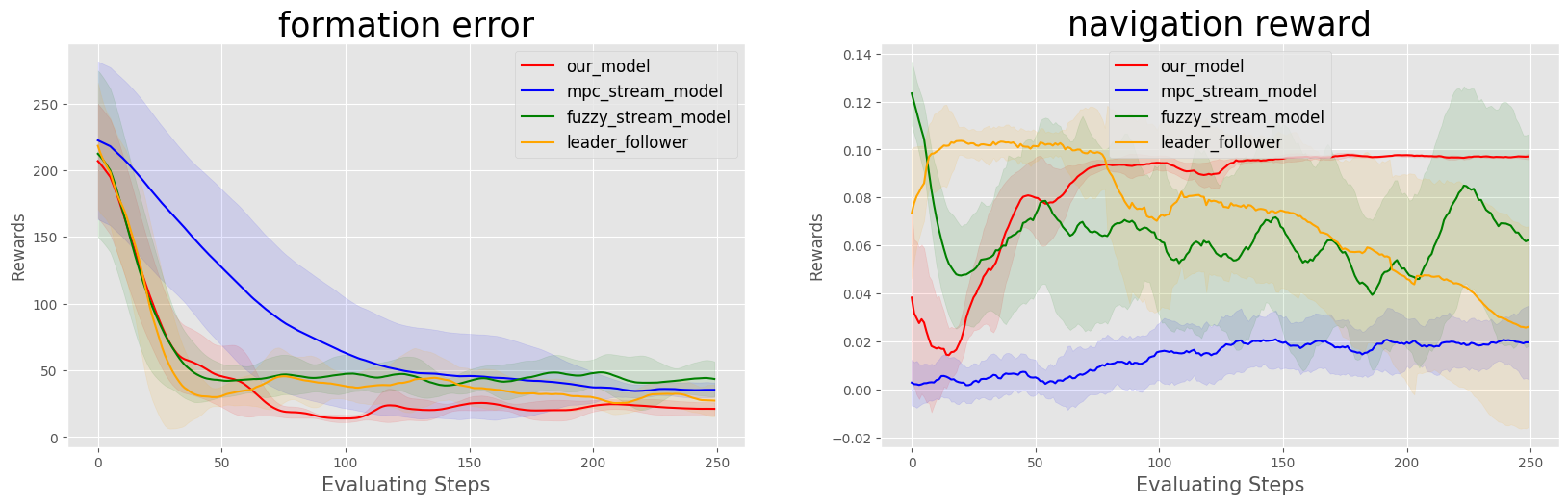}
  \caption{Baseline comparison. Compared with MPC (blue), fuzzy control (green) and leader-follower method (yellow), our model (red) achieves the highest formation accuracy and stability (left side). Meanwhile, our method maintains the most stable and highest navigation speed without any agent crashing obstacles (right side).}
  \label{benchmark}
\end{figure}

Fig.~\ref{benchmark} shows the convergence of the formation error and the navigation reward executed by different controllers. It appears that our method outperforms in formation accuracy, convergence speed and ability to handle multi-tasks. 
Specifically, our method can converge to a lower formation error, which indicates better maintenance of formation during movements. The navigation rewards depict the advantage of DRL scheme when dealing with multi-task problems. The formation is encouraged to accelerate the navigation as the formation error converges, yet the traditional controllers struggle to switch between different behaviors. It is mainly because our method jointly optimizes all reward functions, which enables the agents to successfully learn the priority between formation control and navigation without any prior knowledge. Our method also achieves an on-par success rate of obstacle avoidance with the baselines and achieve higher efficiency of moving to the destinations at the same time.

\subsubsection{Formation Adaptation}

\begin{figure}[tbh]
    \centering
    \subfigure[Change of formation error over time steps in formation adaptation.  In step 144 and 263, two agents are disconnected in sequence. The ideal formation is changed from regular pentagon to square to triangle accordingly.]{
        \includegraphics[height=0.14\textwidth]{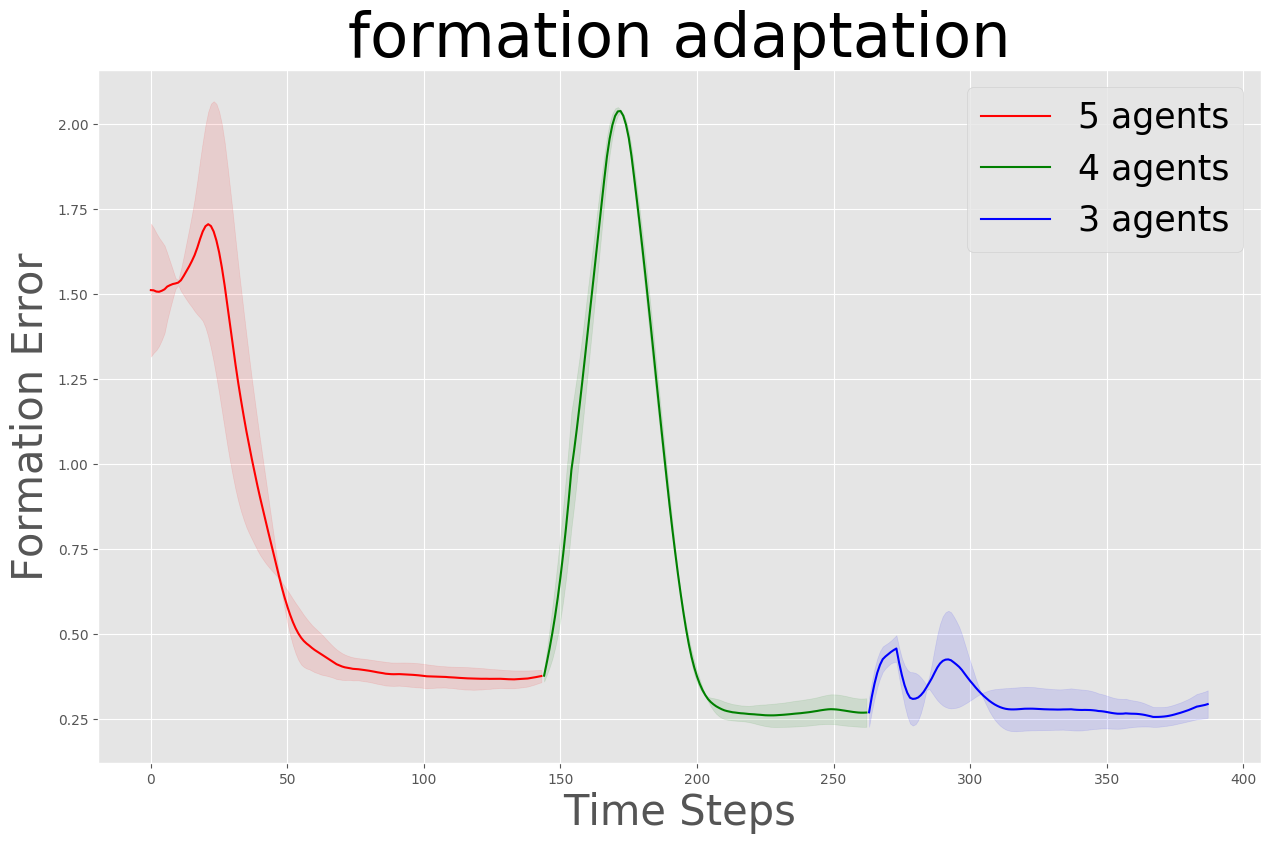}
        \label{fig:ada}
    }
    \subfigure[Promotion of training performance by curriculum learning. The gray line represents the reward curve for training directly in level-4. The colorful lines represent the reward curve for curriculum learning and the difficulty level is changed from level-0 to level-4  in a step-wise process. ]{
        \includegraphics[height=0.14\textwidth]{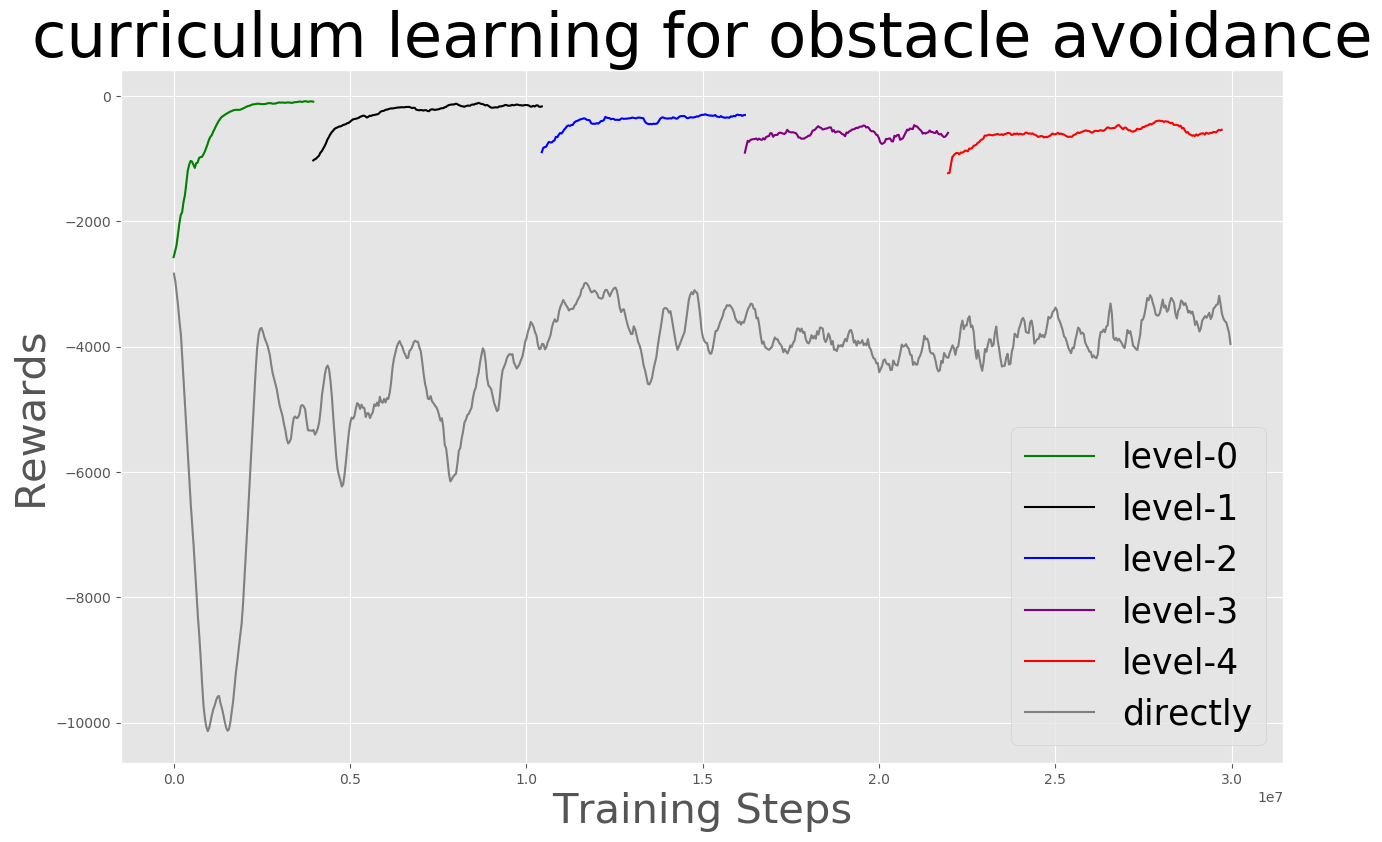}
        \label{fig:curriculum}
    }
    \caption{Results of formation adaptation and curriculum learning.}
    \label{fig:result2} 
\end{figure} 

\begin{figure*}[tbh]
    \centering
    \subfigure[The formation scenario of 3 intelligent vehicle.]{
        \includegraphics[height=0.17\textwidth]{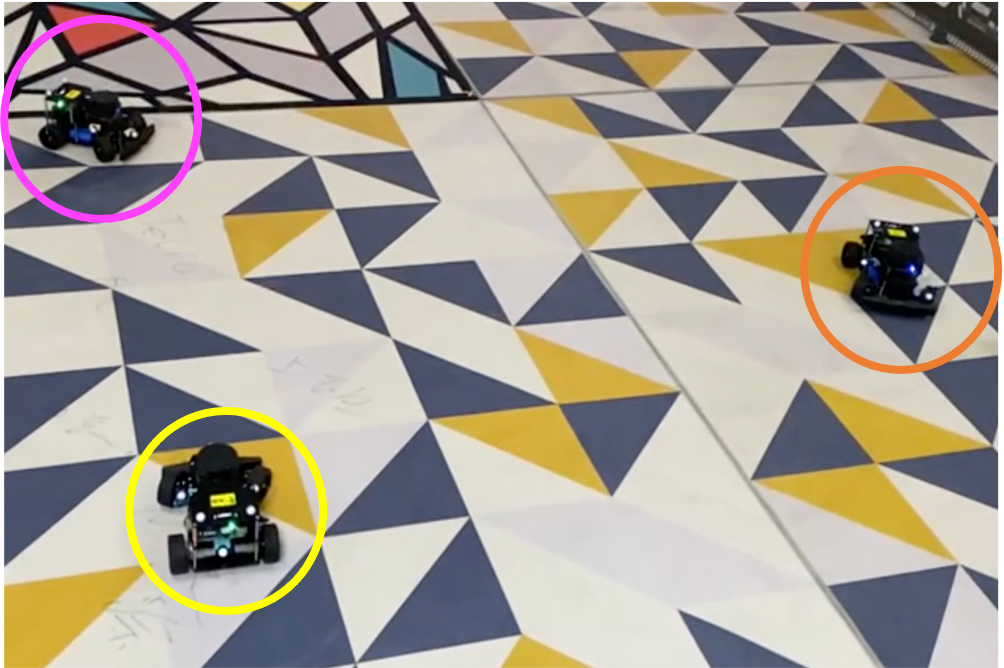}
        \label{platform:overall}
    }
    \subfigure[The overview of the intelligent vehicle equipping with Lidar.]{
        \includegraphics[height=0.17\textwidth]{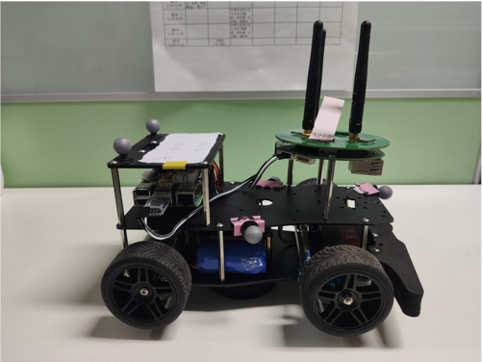}
        \label{platform:agent}
    }
    \subfigure[The chassis and Ackermann-steering gear.]{
        \includegraphics[height=0.17\textwidth]{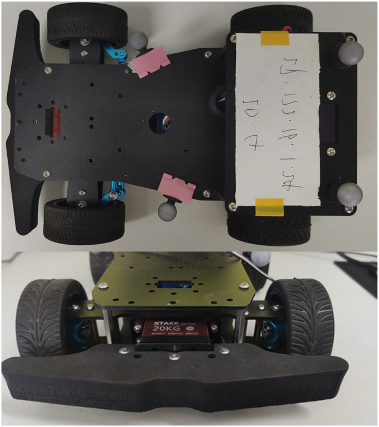}
        \label{platform:steering}
    }
    \subfigure[The OptiTrack motion capture system used for getting the position groundtruth.]{
        \includegraphics[height=0.17\textwidth]{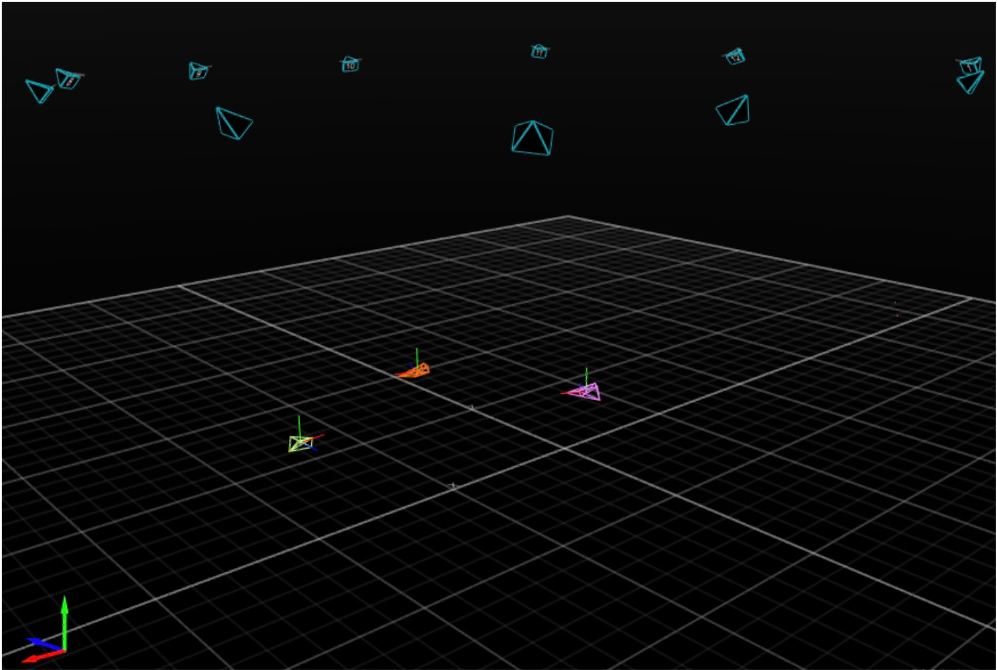}
        \label{platform:optitrack}
    }
    \caption{The hardware platform that is fully consistent with MVE.}
    \label{fig:platform} 
\end{figure*}

The formation policies based on a fixed agent number can be used as teacher models to train a student policy, which is supposed to guide the agents to be reorganized into new formations when some agents are disconnected accidentally. 

In practice, we set the maximum number of agents in the formation to be $5$, and use the formation policies with 3, 4, and 5 agents as the teacher models, respectively. $\alpha$ and $\beta$ in (~\ref{reward3}) are set to be $5$ and $10$. The student model shares the same structure with the teacher model. The sample batch size is set to be $1000$. The Adam optimizer is used with $\beta_1=0.9$, $\beta_2=0.98$, $\epsilon=10^{-3}$. The student model is trained for $5000$ episodes to reach the convergence. The KL divergence is used as loss metric.

We conduct experiments to test the ability of the student policy to guide agents to reorganize their formation when some of them are disconnected. As is shown in Fig.~\ref{fig:ada}, at the beginning of the experiment, 5 agents are guided to form a pentagon formation while moving forward to the destination. In step 144 and 263, we deliberately disconnect 2 agents  in sequence, and the formation will change to square and triangle automatically. Due to changes of the ideal formation topology, the formation error will fluctuate, but new formation will be reorganized and the formation will converge quickly.


\subsubsection{Curriculum Learning for Obstacle Avoidance}

Following~\cite{narvekar2020curriculum}, we set different difficulty levels for obstacle avoidance. 5 levels are set in total from level-0 to level-4, and the corresponding obstacle density is $0/\text{m}^2$, $1\times 10^{-2}/\text{m}^2$, $2\times 10^{-2}/\text{m}^2$, $3\times 10^{-2}/\text{m}^2$, and $5\times 10^{-2}/\text{m}^2$. Initially, no obstacle will be generated in the environment. When the model gets a converged reward under the current level, the difficulty of the task will be enhanced. More obstacles will be generated in the scene to improve the obstacle avoidance ability of the agent cluster. It takes around 6,000,000 steps to train each difficulty level.

As is shown in Fig.~\ref{fig:curriculum}, it is found that if we directly train the policy for formation and obstacle avoidance in complex scenarios with too many obstacles, the model will converge slower due to poor exploration, and the final average reward will be much lower. By curriculum learning, agents can achieve better formation and obstacle avoidance performance in the final level (level-4).  

\subsubsection{Asymmetric Formation}

In addition to formations with regular shapes (e.g. triangle, square, regular polygon), our algorithm can also complete formations with asymmetric topology (e.g. irregular convex polygon). It is found that separated policy is better than shared policy since the relative position of each node is not completely symmetrical. 

\subsection{Hardware Implementation with MVE}
\label{MVE}


The agent in MVE is designed to be an intelligent vehicle with Ackermann-steering. Different from the simplified model in MPE, we consider the constraints of wheelbase and steering angle in practice, which leads to the nonzero turning radius of the vehicle. Given the wheelbase $L$ and the steering angle $\phi$, the turning radius $R$ can be calculated as: $R=\frac{L}{\tan{\phi}}$. Given the the speed of the rear wheels $v_b$, the angular velocity of the orientation angle $\theta$ can be calculated as $\omega=\frac{v_b}{L}\tan{\phi}$. If $L$ is set to be $0$, model will degrade to MPE and we can directly control the vehicle with $v$ and $\omega$. In MVE and hardware implementation, the control variables are changed to $\theta$ and $v_b$. Due to constraints of the hardware control system, we discretize the value of the control variable. The decision frequency is $1$Hz while the control frequency is $100$Hz. At the decision-making stage, the model gives the ideal $v_b$ and $\phi$. At the control stage, the agent will try its best to reach the ideal control variables. The detailed settings can be found at~\url{https://github.com/efc-robot/MultiVehicleEnv}.


\begin{figure}[thpb]
  \centering
  \includegraphics[width=0.8\linewidth]{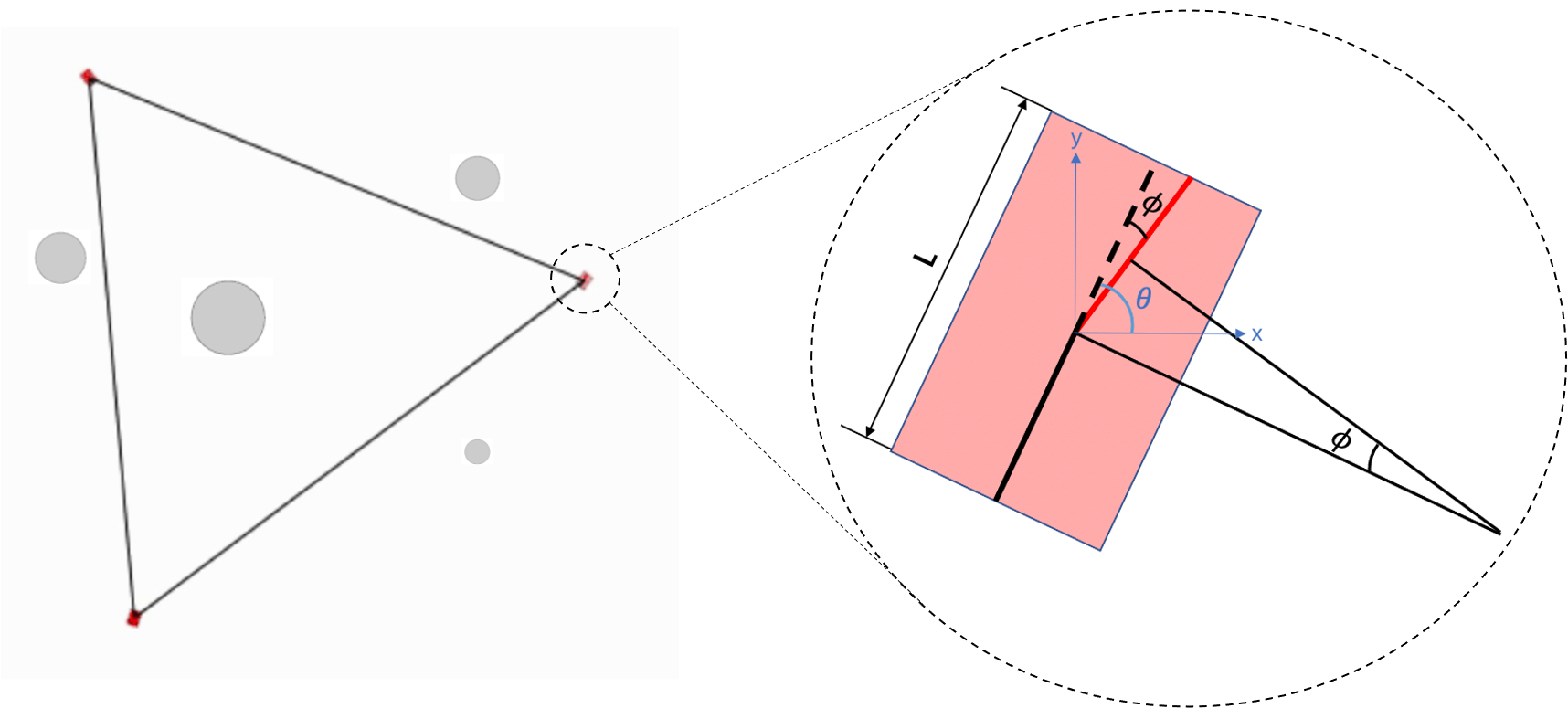}
  \caption{Simulation results in MVE. Left side: The agents (pink rectangles) are forming into a triangle as well as avoiding obstacles (grey circles). Right side: The schematic diagram of Ackermann-steering geometry.   }
  \label{fig:MPE2}
\end{figure}

In hardware implementation, we use scenario as Fig.~\ref{fig:scenario} and Fig.~\ref{platform:overall}. Multiple Ackermann-steering agents are placed in a room randomly. Cylindrical objects with a radius of $0.14$m are placed in the field as obstacles. The intelligent vehicles are as Fig.~\ref{platform:agent}. The wheelbase $L$ is $0.20$m, the overall width is $0.18 $m and the overall length is $0.25 $m, which is fully consistent with the simulation platform as is shown in Fig.~\ref{platform:steering}. The max speed is constrained to $0.361$m/s and the max steering angle is constrained to $0.298\text{rad/s}$.  We use an OptiTrack motion capture system\footnote{ \url{https://www.optitrack.com}}
to get the groundtruth of positions and give relative observations as Fig.~\ref{platform:optitrack}. Agents detect obstacles by Lidars. We 
conduct several hardware experiments with different numbers of agents forming in an environment with obstacles avoidance.
Demos can be found at~\url{https://sgroupresearch.github.io/relativeformation/}.

\section{Conclusion}
In this paper, we develop a MAPPO-based distributed formation and obstacle avoidance method, in which agents only use their local and relative information to make movement decisions. We introduce policy distillation to make the formation system adaptive in case of agents' accidental disconnection. Curriculum learning is also used to simplify the learning process. Our model achieves better performance regarding average formation error, formation convergence rate and success rate of obstacle avoidance. Besides, we also build a new simulation environment and a supporting hardware platform with Ackermann-steering geometry to verify the feasibility of our algorithm. 

For the future work, we will explore large-scale distributed formation methods where agents are not fully connected and can only get the information with the neighbors. Besides, we will also concentrate on developing our self-developed  MVE and corresponding hardware platform in order to solve the sim-to-real problems in MARL algorithm deployments.







\bibliographystyle{IEEEtran}
\bibliography{IEEEabrv,SGroupDefinition,SGroup}

\end{document}